\documentclass[12pt]{article}
\usepackage{epsf}
\usepackage{amsmath,amssymb}
\usepackage{pdfpages}
\usepackage{graphicx}
\usepackage{comment}
\usepackage{tikz}
\usetikzlibrary{positioning}
\usepackage{physics}

\newcommand{\mathhyphen}{\mathchar"712D}

\begin{document}

\newcommand{\rem}[1]{{$\spadesuit$\bf #1$\spadesuit$}}

\renewcommand{\thefootnote}{\fnsymbol{footnote}}
\setcounter{footnote}{0}

\begin{titlepage}

\def\thefootnote{\fnsymbol{footnote}}

\begin{center}
\hfill TU-1125\\
\hfill May, 2021\\

\vskip .75in

{\Large \bf

  Hidden Dark Matter from Starobinsky Inflation

}

\vskip .5in

{\large
  Qiang Li$^{(a)}$, Takeo Moroi$^{(a)}$, Kazunori Nakayama$^{(a)}$
  and Wen Yin$^{(b, a)}$
}

\vskip 0.5in

$^{(a)}${\em
Department of Physics, The University of Tokyo, Tokyo 113-0033, Japan
}

\vskip 0.2in

$^{(b)}${\em 
Department of Physics, Tohoku University, Sendai 980-8578, Japan
}

\end{center}
\vskip .5in

\begin{abstract}

  The Starobinsky inflation model is one of the simplest inflation
  models that is consistent with the cosmic microwave background
  observations. In order to explain dark matter of the universe, we
  consider a minimal extension of the Starobinsky inflation model with
  introducing the dark sector which communicates with the visible
  sector only via the gravitational interaction.  In Starobinsky
  inflation model, a sizable amount of dark-sector particle may be
  produced by the inflaton decay.  Thus, a scalar, a fermion or a
  vector boson in the dark sector may become dark matter.  We pay
  particular attention to the case with dark non-Abelian gauge
  interaction to make a dark glueball a dark matter candidate.  In the
  minimal setup, we show that it is difficult to explain the observed
  dark matter abundance without conflicting observational constraints
  on the coldness and the self-interaction of dark matter.  We propose
  scenarios in which the dark glueball, as well as other dark-sector
  particles, from the inflaton decay become viable dark matter
  candidates.  We also discuss possibilities to test such scenarios.

\end{abstract}

\end{titlepage}

\renewcommand{\thepage}{\arabic{page}}
\setcounter{page}{1}
\renewcommand{\thefootnote}{\#\arabic{footnote}}
\setcounter{footnote}{0}
\renewcommand{\theequation}{\thesection.\arabic{equation}}

\tableofcontents

\section{Introduction}
\label{sec:intro}
\setcounter{equation}{0}

A simple model of inflation consistent with the cosmic microwave
background (CMB) observation by the Planck
satellite~\cite{Akrami:2018odb} is the so-called Starobinsky inflation
model~\cite{Starobinsky:1980te}. In the Starobinsky model, the $R^2$
term is added to the Einstein-Hilbert action where $R$ denotes the
Ricci scalar. This model contains a scalar degree of freedom, called
scalaron, and it can be regarded as the inflaton if the coefficient of
the $R^2$ term takes an appropriate value.  One interesting feature of
the Starobinsky model is that the inflaton universally couples to the
Standard Model sector as well as other sectors through the trace of
the energy-momentum tensor, and hence the reheating temperature is
predictable.  In order to explain the dark matter (DM) of the
universe, we may consider models with the dark sector containing DM,
which is sequestered from the visible sector and communicates with the
visible sector only via the gravitational interaction.  Then, it is
also possible to predict the inflaton decay rate into the dark sector.

In this paper we consider a simple dark sector extension to the
Starobinsky inflation in order to explain DM.  Such a scenario has
been discussed in \cite{Gorbunov:2010bn, Gorbunov:2012ns,
  Bernal:2020qyu}. It has been shown that an addition of massive
scalar or fermion, as well as $SU(N)$ gauge boson may be enough to
explain the correct DM abundance through the inflaton decay.  We
revisit this issue.  In particular, we pay particular attention to the
case of dark (pure) non-Abelian gauge sector.  A characteristic
property of the dark glueball DM is that it has a sizable
self-interaction cross section.  In addition, we also consider the
cases that the DM candidate is a scalar boson, a fermion, and a
massive Abelian gauge boson which was not discussed in the previous
studies.

Let us mention several relevant works.  In the past, the
self-interacting DM was proposed to alter the density
perturbation~\cite{Carlson:1992fn} where the entropy of the dark
sector is given by hand (see also \cite{Dolgov:1980uu,
  Dolgov:2017ujf}).  Recently such kinds of scenarios have become
attracting attentions again \cite{Hochberg:2014kqa, Kuflik:2015isi,
  Pappadopulo:2016pkp, Farina:2016llk} since they may explain the
small-scale crisis and alleviate the $S_8$
tension~\cite{Spergel:1999mh, deBlok:2009sp, BoylanKolchin:2011de,
  Salucci:2018hqu, Heimersheim:2020aoc}.  The DM production in this
context has also been popularly studied.  For instance the DM can be
produced via the coupling to SM
particles~\cite{Bernal:2015ova,Hochberg:2015vrg, Chu:2017msm}, or via
the inflaton decay by assuming the vanishing coupling to the SM
particles~\cite{Bernal:2018hjm}. (See also the DM production from
inflaton decay in a general context for heavy DM candidates
\cite{Moroi:1994rs, Kawasaki:1995cy, Moroi:1999zb, Jeong:2011sg,
  Ellis:2015jpg, Kaneta:2019zgw} and light sub-keV DM candidates
\cite{Moroi:2020has, Moroi:2020bkq}.)  Our scenario may be regarded as
a concrete model of self-interacting DM produced by the inflaton
decay.  In particular, we consider the Starobinsky inflation, which is
motivated from the current CMB observation, and assume a dark sector
containing a DM candidate, which automatically suppresses the coupling
to the SM particles.  We pay particular attention to the case with a
dark $SU(N)$ gauge group whose dark glueball becomes the DM
candidate \cite{Faraggi:2000pv, Feng:2011ik, Boddy:2014yra,
  Soni:2016gzf, Gross:2020zam}.  Interestingly, the
inflaton coupling to the dark sector is determined, and hence we
have rigid predictions.  We find that it is difficult to explain the
present DM in the minimal setup since the constraint on the DM
self-interaction cross section is in tension with the prediction DM
abundance.  We also discuss several loopholes to avoid such
constraints.

This paper is organized as follows.  In Section \ref{sec:model}, we
give an overview of the model of our interest.  In Section
\ref{sec:minimal}, we study the possibility of the dark glueball DM in
the minimal setup.  We will see that the dark glueball DM is hardly
realized in the minimal setup.  In Section \ref{sec:glueballdm}, we
propose scenarios to make the dark glueball a viable candidate of the
DM.  In Section \ref{sec:others}, we discuss the cases that the dark
sector contains a scalar boson, a fermion, or a massive Abelian gauge
boson as a candidate of the DM.  In Section \ref{sec:conclusions}, we
summarize our results.

\section{Model}
\label{sec:model}
\setcounter{equation}{0}

\subsection{Starobinsky inflation model}

In this section, we summarize basic properties of the Starobinsky
inflation which are necessary for our analysis.  We start with the
action in the Jordan frame for the Starobinsky inflation \cite{Starobinsky:1980te}:
\begin{align}
  S = S_{\rm grav} + S_{\rm vis} + S_{\rm dark},
  \label{S_starobinsky}
\end{align}
with 
\begin{align}
  S_{\rm grav} \equiv 
  -\frac{1}{2} M_{\rm Pl}^2
  \int d^4 x \sqrt{-\hat{g}}
  \left( \hat{R} - \frac{1}{6\mu^2} \hat{R}^2 \right),
  \label{Sgrav}
\end{align}
where $M_{\rm Pl}\simeq 2.4\times 10^{18}\ {\rm GeV}$ is the reduced
Planck scale, $R$ denotes the Ricci scalar, and $S_{\rm vis}$ and
$S_{\rm dark}$ are actions for the visible sector (which contains the
SM particles) and dark sector (which contains DM),
respectively.\footnote
{We adopt the sign convention of the metric of $g_{\mu\nu}\sim
  (+,-,-,-)$.  The Ricci scalar during inflation is given by $R\simeq
  -12H^2$, with $H$ being the expansion rate.}
Here and hereafter, the ``hat'' is for quantities in the Jordan frame.

The action $S_{\rm grav}$ is equivalent to the following one (for the
review of so-called $f(R)$ gravity, see, for example,
\cite{DeFelice:2010aj}):
\begin{align}
  S_{\rm grav} \equiv
  -\frac{1}{2} M_{\rm Pl}^2
  \int d^4 x \sqrt{-\hat{g}}
  \left[ \varphi \hat{R} + \frac{3}{2} \mu^2 (\varphi -1)^2 \right].
  \label{Eeq}
\end{align}
Regarding $\varphi$ as an auxiliary field, the Eular-Lagrange equation
gives $\varphi=1-\frac{1}{3\mu^2}\hat{R}$, with which the above action
becomes equal to the original one given in Eq.\ \eqref{Sgrav}.  In the
following, we use the action given in Eq.\ \eqref{Eeq}, which is more
convenient for our discussion.

With the action $S_{\rm grav}$ given above, there exists a physical
scalar degree of freedom which is often called scalaron; the scalaron
plays the role of the inflaton in the Starobinsky inflation model.  To
see this, it is convenient to work in the Einstein frame.  The
Einstein-Hilbert action can be obtained with the following Weyl
transformation:
\begin{align}
  g_{\mu\nu} = \varphi \hat{g}_{\mu\nu},
\end{align}
with which $\hat{R}$ and the Ricci scalar in the Einstein frame $R$ are
related as 
\begin{align}
  \hat{R} = \varphi
  \left[ R + 3 \frac{\Box\, \varphi}{\varphi} 
    -\frac{9}{2} g^{\mu\nu} \partial_\mu \varphi \partial_\nu \varphi
    \right].
\end{align}
We can see that there exists a scalar degree of freedom; the
canonically normalized scalar field $\phi$ is related to $\varphi$ as
\begin{align}
  \varphi = \exp \left( \sqrt{\frac{2}{3}} \frac{\phi}{M_{\rm Pl}}\right).
\end{align}
In terms of $\phi$, the Einstein frame action for the gravity sector
is given by
\begin{align}
  S_{\rm grav} = 
  \int d^4 x \sqrt{-g}
  \left[
    -\frac{1}{2} M_{\rm Pl}^2 R
    + \frac{1}{2} g^{\mu\nu} \partial_\mu \phi \partial_\nu \phi - V(\phi)
  \right],
\end{align}
where the potential of the scalaron is given by
\begin{align}
  V(\phi) \equiv \frac{3}{4} \mu^2 M_{\rm Pl}^2
  \left[
    1 - \exp \left( -\sqrt{\frac{2}{3}} \frac{\phi}{M_{\rm Pl}}\right)
    \right]^2.
  \label{Vphi}
\end{align}
The minimum of the potential of $\phi$ is chosen to be $\phi=0$;
expanding the potential around the minimum, we can find
\begin{align}
  V(\phi) \simeq \frac{1}{2} \mu^2 \phi^2 + O(\phi^3),
\end{align}
and hence $\mu$ can be regarded as the mass of $\phi$ in the vacuum:
\begin{align}
  m_\phi = \mu.
\end{align}
As one can see from Eq.\ \eqref{Vphi}, the potential of $\phi$ becomes
flat when $\phi\rightarrow\infty$.  Thus, if the initial amplitude of
$\phi$ is large enough, the slow-roll inflation can occur.

Regarding $V(\phi)$ as the inflaton potential, the curvature
perturbation amplitude $A_s$, the scalar spectral index $n_s$, and the
tensor-to-scalar ratio $r$ are evaluated as
\begin{align}
  A_s (k) \simeq &\, \frac{1}{24\pi^2} \frac{\mu^2}{M_{\rm Pl}^2} N_e^2,
  \\
  n_s (k) -1 \simeq &\, - \frac{2}{N_e},
  \\
  r (k) \simeq &\, \frac{12}{N_e^2},
\end{align}
where $N_e$ is the $e$-folding number for the epoch at which the mode
$k$ (with $k$ being the present physical wave number) exits the
horizon, and is related to the inflaton amplitude during inflation as
\begin{align}
  N_e \simeq \frac{3}{4}
  \exp \left( \sqrt{\frac{2}{3}} \frac{\phi}{M_{\rm Pl}}\right).
\end{align}
Taking $N_e (k=0.05\ {\rm Mpc}^{-1})\simeq 50-60$, the observed value of $A_s\simeq 2.1\times
10^{-9}$ \cite{Aghanim:2018eyx} is realized with $\mu\simeq 3\times
10^{13}\ {\rm GeV}$.  The scalar spectral index becomes
consistent with the observed value while the tensor-to-scalar ratio is
well below the current upper bound \cite{Akrami:2018odb}.

\subsection{Reheating}

A brief overview the cosmic evolution is as follows.  Assuming that
the initial amplitude of $\phi$ is large enough, the Starobinsky
inflation occurs, during which $\phi$ slowly moves towards the minimum
of the potential.  When the inflaton amplitude becomes $\sim M_{\rm
  Pl}$, the slow-roll condition breaks down and the inflation ends.
Then, the inflaton starts to oscillate around the potential minimum;
soon after the start of the oscillation and well before the decay, the
energy density of the inflaton scales as $a^{-3}$ (with $a$ being the
cosmic scale factor).  When the expansion rate of the universe becomes
comparable to the total decay rate of the inflaton, denoted as
$\Gamma_\phi$, the inflaton decays effectively and the universe is
reheated.

In order to discuss the detail of the reheating process, we consider
the interaction of the inflaton with the matter particles.  The
interaction terms of the inflaton in the Einstein frame show up after
the Weyl transformation, and are expressed in the following form:
\begin{align}
  {\cal L}_{\rm int} = \frac{1}{\sqrt{6} M_{\rm Pl}} \phi T^\mu_\mu,
  \label{Lint}
\end{align}
where $T_{\mu\nu}$ is the energy momentum tensor:
\begin{align}
  T_{\mu\nu} = \frac{2}{\sqrt{-g}} 
  \frac{\delta (S_{\rm vis} + S_{\rm dark})}{\delta g^{\mu\nu}}.
\end{align}
With Eq.\ \eqref{Lint}, we can calculate the decay rate of the
inflaton.  The partial decay rates depend strongly on the properties
of the final-state particles.  Below we list them for the cases of a
final state scalar, a fermion and a vector boson.
\begin{itemize}
\item For a scalar field $\chi$, we consider the following
  Lagrangian:\footnote
  {Here, we neglect the interaction of the scalar field because
    we are interested in the two body decay process
    $\phi\rightarrow\chi\chi$.}  
  \begin{align}
    {\cal L}_\chi =
    \frac{1}{2} \hat g^{\mu\nu} \partial_\mu \chi \partial_\nu \chi
    - \frac{1}{2} m_\chi^2 \chi^2
    + \frac{1}{2}\xi_\chi \hat R \chi^2,  \label{L_chi}
  \end{align}
  where we introduce the non-minimal coupling constant $\xi_\chi$.
  Here we assume that the scalar DM is charged under a $Z_2$
  symmetry to forbid the interaction $\hat R \chi$. We will come back
  to remove this assumption later.  Then, the decay rate is given by
  \begin{align}
    \Gamma_{\phi\rightarrow\chi\chi} =
    \frac{(1-6\xi_\chi)^2}{192\pi} \frac{m_\phi^3}{M_{\rm Pl}^2},
    \label{Gamma_chichi}
  \end{align}
  where we assume that $\chi$ is much lighter than $\phi$.
\item For a Dirac fermion $\psi$ with 
  \begin{align}
    {\cal L}_\psi = i \bar{\psi} \gamma^\mu \partial_\mu \psi
    - m_\psi \bar{\psi} \psi,
  \end{align}
  the decay rate is given by
  \begin{align}
    \Gamma_{\phi\rightarrow\bar{\psi}\psi} =
    \frac{1}{48\pi} \frac{m_\psi^2 m_\phi}{M_{\rm Pl}^2}.
    \label{Gamma_ff}
  \end{align}
\item We may also consider the decay of $\phi$ into a pair of gauge
  bosons in association with a gauge group $G$.  At the tree level,
  the partial decay rate $\Gamma_{\phi\rightarrow G_\mu G_\mu}$ (with
  $G_\mu$ denoting the gauge boson) vanishes because the energy
  momentum tensor of the gauge boson is traceless.  With taking into
  account quantum effects, however, this is not the case because
  \begin{align}
    T_\mu^{(G)\mu} = \frac{\beta_G (\alpha_G)}{4\alpha_G}
    G_{\mu\nu}^{(a)} G^{(a)\mu\nu},
  \end{align}
  where $\alpha_G$ is the gauge coupling constant, $\beta_G$ is the
  $\beta$-function of $\alpha_G$, and $G_{\mu\nu}^{(a)}$ is the field
  strength tensor.  We use the 1-loop $\beta$-function:
  \begin{align}
    \beta_G^{\rm (1\mathhyphen loop)} (\alpha_G) =
    \frac{b_G}{2\pi} \alpha_G^2,
  \end{align}
  where $b_G$ is a constant with which the coupling constant at two
  different renormalization scales, $Q$ and $Q_0$, are related as
  \begin{align}
    \alpha_G (Q) =
    \left[ \alpha_G^{-1} (Q_0) - \frac{b_G}{2\pi} \log (Q/Q_0) \right]^{-1}.
  \end{align}
  Then, the decay rate is given by \cite{Gorbunov:2012ns}
  \begin{align}
    \Gamma_{\phi\rightarrow G_\mu G_\mu} =
    \frac{D_G b_G^2 \alpha_G^2}{1536 \pi^3} \frac{\mu^3}{M_{\rm Pl}^2},
    \label{Gamma(gauge)}
  \end{align}
  where $D_G$ is the number of gauge bosons (i.e., dimension of the
  group $G$).
\item One may consider a massive Abelian gauge boson $V_\mu$ (with its
  field strength denoted as $V_{\mu\nu}$) with the Lagrangian
  \begin{align}
    {\cal L}_V =
    -\frac{1}{4} V_{\mu\nu} V^{\mu\nu}
    + \frac{1}{2} m_V^2 V_\mu V^\mu.
  \end{align}
  When the mass of $V_\mu$ is much lighter than the inflaton mass, the
  production of the longitudinal mode dominates and the decay rate is
  given by
  \begin{align}
    \Gamma_{\phi\rightarrow V_\mu V_\mu} =
    \frac{1}{192\pi} \frac{m_\phi^3}{M_{\rm Pl}^2}.
    \label{Gamma_VV}
  \end{align}
  The mass of $V_\mu$ may arise from the spontaneous breaking of the
  gauge symmetry due to the vacuum expectation value of a scalar field
  charged under the $U(1)$ symmetry.  For simplicity, here and
  hereafter we consider the case that the Higgs mode in association
  with the breaking of the $U(1)$ symmetry is much heavier than the
  inflaton.  Such a case is realized when $v\rightarrow\infty$ and
  $g_{U(1)}\rightarrow 0$ (where $v$ is the vacuum expectation value
  of the scalar field responsible for the $U(1)$ symmetry breaking and
  $g_{U(1)}$ the gauge coupling constant) while keeping the product
  $g_{U(1)}v$ finite.  If $g_{U(1)}$ is exactly equal to $0$, the
  decay process should be understood as the production of the
  Nambu-Goldstone (NG) mode.  When $v\rightarrow\infty$, the Higgs
  mode should decouple from the low energy spectrum and we can
  consider an effective field theory containing only the inflaton and
  the NG mode.  In such a case, using the fact that the non-minimal
  coupling of the NG scalar is forbidden by the shift symmetry, we can
  see that the decay rate given in Eq.\ \eqref{Gamma_VV}
  parametrically agrees with the decay rate to the NG mode (see
  Eq.\ \eqref{Gamma_chichi}).
\end{itemize}

In order for a viable thermal history, the inflaton should dominantly
decay into visible sector fields.  The inflaton dominantly decays into
scalar fields unless the scalars are all conformally coupled (i.e.,
$\xi_\chi\simeq \frac{1}{6}$).  If the decay processes into the
scalars are suppressed, the inflaton may also decay into gauge fields.\footnote{There are also decay processes to 
SM fermion pairs and a Higgs boson via the trace anomaly. The three-body decay rates are suppressed by the phase spaces and/or loop factors. 
}
Here, we evaluate the decay rate of the inflaton into the
visible-sector field as
\begin{align}
  \Gamma_{\phi\rightarrow {\rm vis}} =
  \frac{1}{192 \pi}
  \left[ 
    4 (1 - 6 \xi_h)^2 
    + \sum_{A=1,2,3}
    \frac{D_{A} b_A^2 \alpha_A^2}{8 \pi^2}
    \right] \frac{m_\phi^3}{M_{\rm Pl}^2},
  \label{Gamma(vis)}
\end{align}
by taking account of the SM Higgs field and gauge bosons as final
states, where $\xi_h$ is the SM Higgs non-minimal coupling to the
Ricci scalar in the form of (\ref{L_chi}).  In addition, $A=1$, $2$,
and $3$ are contributions of $U(1)_Y$, $SU(2)_L$, and $SU(3)_C$ gauge
groups, respectively, with $D_1=1$, $D_2=3$, and $D_3=8$.  Hereafter,
for simplicity, we assume that the $\beta$-functions of the SM gauge
coupling constants are determined only by the SM fields (at least at
the $1$-loop level), and take $b_1=\frac{41}{6}$, $b_2=-\frac{19}{6}$,
and $b_3=-7$.

We can relate $\Gamma_{\phi\rightarrow {\rm vis}}$ with the reheating
temperature after the inflaton decay.  In our discussion, we consider
the case that the inflaton dominantly decays into visible-sector
fields.  Then, the reheating temperature, which is the visible-sector
temperature at the end of the inflaton decay, is evaluated as

\begin{align}
  T_{\rm vis} (t_{\rm R}) \equiv
  \left(\frac{g_\star (t_{\rm R}) \pi^2 }{90}\right)^{-1/4} 
  \sqrt{ M_{\rm Pl} \Gamma_{\phi \to {\rm vis}} },
\end{align}
where $t_{\rm R}$ is the cosmic time at the end of the reheating, and
$g_\star$ is the relativistic degrees of freedom of the SM.  Assuming
that $\phi \to h^\dagger h $ is the dominant decay channel of $\phi$, with $h$ being the SM Higgs field,
$g_\star (t_{\rm R})=106.75$ and $\mu =3\times 10^{13}\ {\rm GeV}$, we obtain
\begin{align}
  T_{\rm vis} (t_{\rm R}) \simeq 5\times 10^9\ {\rm GeV} \times
  |1-6\xi_{h}|.
  \label{rehSM}
\end{align}

Here let us comment on the ``gravitational particle production'' in the Starobinsky model.
The gravitational particle production refers to the creation of fields that do not have any direct interaction to the inflaton at the Lagrangian level. In this sense, the inflaton decay in the Starobinsky model may be regarded as a result of gravitational production, since the inflaton sector and the matter sector are decoupled in the Jordan frame (\ref{S_starobinsky}).
The gravitational particle production during the reheating phase\footnote{
	Gravitational particle production often means the particle creation during inflation or at the transition epoch from de Sitter to the reheating phase~\cite{Ford:1986sy,Chung:1998zb,Chung:2001cb,Chung:2011ck}. In particular, the vector DM production in this context has been discussed in Refs.~\cite{Graham:2015rva,Ema:2019yrd,Ahmed:2020fhc,Kolb:2020fwh,Gross:2020zam}	
}  has been discussed in Refs.~\cite{Ema:2015dka,Ema:2016hlw,Ema:2018ucl,Ema:2019yrd}. It is shown that the gravitational particle production picture in the Jordan frame gives the same result as the inflaton decay picture in the Einstein frame~\cite{Arbuzova:2011fu,Ema:2016hlw}.

\section{Mass Density of Dark Glueball: Minimal Case}
\label{sec:minimal}
\setcounter{equation}{0}

First, we discuss the possibility that the dark glueball in the dark
sector becomes DM \cite{Faraggi:2000pv, Feng:2011ik, Boddy:2014yra,
  Soni:2016gzf, Gross:2020zam}.  In this Section, we consider the
minimal case that the visible sector contains the SM particles while
the dark sector contains an $SU(N)$ pure Yang-Mills gauge sector.
In this case, after the inflation, the inflaton decays into the SM
particles as well as the dark gluon.  Then, after the confinement of
$SU(N)$, which is expected to occur when the temperature in the dark
sector becomes lower than 
the dynamical scale, the dark glueballs are
formed.  We denote the lightest dark glueball as $X$.  We are
interested in the production of the dark glueball from the inflaton
decay, and we assume that dark glueball mass $m_X$ is smaller than the
inflaton mass.

\subsection{Abundance of dark glueball}

\begin{figure}[t]
  \centering
  \includegraphics[width=0.75\textwidth]{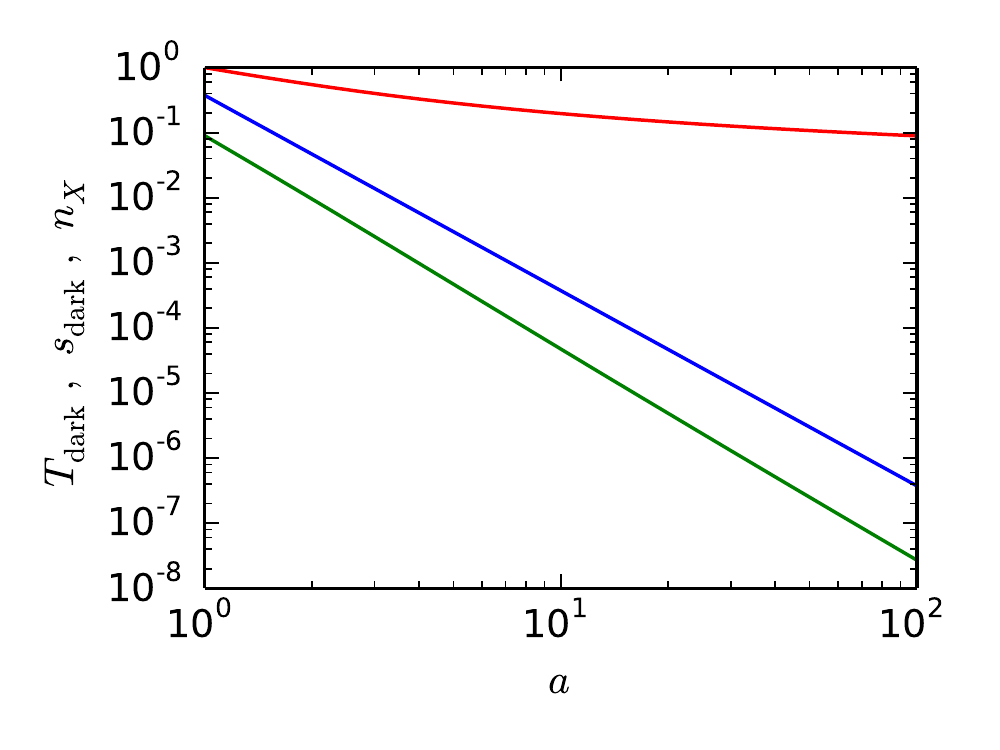}
  \caption{The red, blue, and green lines show $T_{\rm dark}$ (normalized by
    $\Lambda$), $s_{\rm dark}$ (normalized by $\Lambda^3$), and
    $n_{\rm dark}$ (normalized by $\Lambda^3$) as functions of the
    scale factor, assuming the chemical equilibrium; the scale
    factor is normalized as $a|_{T_{\rm dark}=\Lambda}=1$.}
  \label{fig:stn}
\end{figure}

With the cosmic expansion, the universe evolves as follows:
\begin{itemize}
\item After inflation, the dark gluon as well as the SM particles are
  produced by the inflaton decay.  The partial decay rate of the
  inflaton into the dark gluon pair is given by
  Eq.\ \eqref{Gamma(gauge)} with the $\beta$-function coefficient:
  \begin{align}
    b_{\rm dark} = - \frac{11}{3} N.
  \end{align}
  The gauge coupling constant at the renormalization scale $Q$ is
  evaluated as
  \begin{align}
    \alpha_{\rm dark} (Q) = -\frac{2\pi}{b_{\rm dark} \log (Q/\Lambda)},
  \end{align}
  where $\Lambda$ is the dynamical scale of the dark $SU(N)$ gauge
  theory.  In addition, 
  \begin{align}
    D_{\rm dark} \equiv N^2 -1.
  \end{align}
\item When $\rho_{\rm dark}\gg\Lambda^4$, the dark sector is described
  as a relativistic matter consisting of dark gluons.  In such an era,
  assuming that the SM particles are also relativistic, the evolutions
  of the energy densities of $\phi$, the visible sector, and the dark
  sector, denoted as $\rho_\phi$, $\rho_{\rm vis}$, and $\rho_{\rm
    dark}$, respectively, are governed by
  \begin{align}
    \dot{\rho}_\phi = &\,
    - 3 H \rho_\phi 
    - (\Gamma_{\phi\rightarrow {\rm vis}} + \Gamma_{\phi\rightarrow {\rm dark}})
    \rho_\phi,
    \label{rhodot_phi}
    \\
    \dot{\rho}_{\rm vis} = &\,
    - 4 H \rho_{\rm vis} + \Gamma_{\phi\rightarrow {\rm vis}} \rho_\phi,
    \label{rhodot_vis}
    \\
    \dot{\rho}_{\rm dark} = &\,
    - 4 H \rho_{\rm dark} + \Gamma_{\phi\rightarrow {\rm dark}} \rho_\phi,
    \label{rhodot_dark}
  \end{align}
  where the ``dot'' denotes the derivative with respect to time, and
  the expansion rate $H$ is given by
  \begin{align}
    H = \sqrt{\frac{\rho_{\rm tot}}{3 M_{\rm Pl}}},
  \end{align}
  with $\rho_{\rm tot}\equiv\rho_\phi+\rho_{\rm vis}+\rho_{\rm dark}$.
  The temperature of the dark sector in such an era is related to
  $\rho_{\rm dark}$ as
  \begin{align}
    \rho_{\rm dark} = D_{\rm dark} \frac{\pi^2}{30} T_{\rm dark}^4,
  \end{align}
  while the entropy density in the dark sector is given by
  \begin{align}
    s_{\rm dark} = D_{\rm dark} \frac{\pi^2}{45} T_{\rm dark}^3.
  \end{align}
  After the completion of the inflaton decay, $s_{\rm vis}$ (which is
  the entropy density of the visible sector) and $s_{\rm dark}$ are
  both proportional to $a^{-3}$.  The dark-sector temperature at the
  time of the reheating is estimated as
  \begin{align}
    T_{\rm dark} (t_{\rm R}) \simeq 
    4\times 10^9\ {\rm GeV} \times \sqrt{\alpha_G b_G} |1-6\xi_h|^{1/2},
    \label{rehdark}
  \end{align}
  where we have assumed that the inflaton dominantly decays into the
  Higgs pair.
  Notice that, even after the confinement of dark $SU(N)$, $s_{\rm
    dark}\propto a^{-3}$, assuming that there is no extra entropy production like a first order phase
  transition to the confinement phase.
\item When the dark-sector temperature $T_{\rm dark}$ becomes lower
  than $\sim\Lambda$, we expect that the confinement of the dark
  $SU(N)$ occurs.  Then, the dark sector contains only the dark
  glueball degrees of freedom.  Because there is no symmetry
  forbidding $2\leftrightarrow 3$ (and other number changing)
  processes of dark glueballs, the dark glueballs are in chemical
  equilibrium as far as the scattering rate of the number changing
  processes is larger than the expansion rate of the universe.
  Assuming that the thermal bath of the dark sector is dominated by
  the lightest dark glueball, and that the dark glueball is in
  chemical equilibrium, the energy density, pressure, and entropy
  density of the dark sector are given by
  \begin{align}
    \rho_{\rm dark} = &\,
    \frac{1}{2\pi^2} \int_0^\infty dk
    \frac{k^2 E_k}{e^{E_k/T_{\rm dark}-1}},
    \label{rhodark(NR)}
    \\
    p_{\rm dark} = &\,
    \frac{1}{2\pi^2} \int_0^\infty dk
    \frac{k^4}{3E_k (e^{E_k/T_{\rm dark}-1})},
    \label{pdark(NR)}
    \\
    s_{\rm dark} = &\, \frac{\rho_{\rm dark}+p_{\rm dark}}{T_{\rm dark}},
    \label{sdark(NR)}
  \end{align}
  respectively, where $E_k\equiv\sqrt{k^2+m_X^2}$, with $m_X$ being
  the mass of dark glueball.  The number density of the dark glueball
  is given by
  \begin{align}
    n_X = \frac{1}{2\pi^2} \int_0^\infty dk
    \frac{k^2}{e^{E_k/T_{\rm dark}-1}}.
    \label{n(NR)}
  \end{align}
  Because of the conservation of the entropy in the comoving volume,
  the dark-sector temperature does not decrease rapidly after the
  confinement of the dark $SU(N)$ as far as the chemical equilibrium
  is maintained in the dark sector \cite{Carlson:1992fn}.  When the
  chemical equilibrium is maintained in the dark sector, $n_X$
  decreases slightly slower than $a^{-3}$, while the temperature
  scales as $T_{\rm dark}\propto 1/\ln a$.  In Fig.\ \ref{fig:stn}, we plot
  $T_{\rm dark}$ (normalized by $\Lambda$), $s_{\rm dark}$ (normalized by
  $\Lambda^3$), and $n_{\rm dark}$ (normalized by $\Lambda^3$) as
  functions of the scale factor, assuming the chemical equilibrium; in
  the figure, the scale factor is normalized as
  $a|_{T_{\rm dark}=\Lambda}=1$.
\item When the rate of the number changing processes becomes smaller
  then the expansion rate, the number of the glueball in the comoving
  volume is fixed.  Even in such an era, the $2$ to $2$ elastic
  scattering may be effective; we will consider the effect of the
  elastic scattering later.
\end{itemize}
In our analysis, we assume the adiabatic expansion of the universe
after the reheating.  This is the case for $N=2$; the confinement
phase transition is second order for $N=2$, while it is first order
for $N\geq 3$ \cite{Lucini:2005vg}.\footnote
{This argument may depend on the value of the strong CP phase of the
  dark sector.}
In the following discussion, we take $N=2$. 
A Comment on the case of $N\geq 3$ will be given at the end of this Section.

In order to estimate the present mass density of the dark glueball, we
first solve Eqs.\ \eqref{rhodot_phi} -- \eqref{rhodot_dark}.  From
Eqs.\ \eqref{rhodot_vis} and \eqref{rhodot_dark}, the ratio of the
energy densities of the visible and dark sectors are given by
\begin{align}
  \left. \frac{\rho_{\rm dark}}{\rho_{\rm vis}} \right|_{T_{\rm dark}\gg\Lambda}
  =
  \frac{\Gamma_{\phi\rightarrow{\rm dark}}}{\Gamma_{\phi\rightarrow{\rm vis}}}.
\end{align}
In addition, we use the fact that the entropy ratio $s_{\rm
  dark}/s_{\rm vis}$ is a constant of time; in our analysis, we
calculate the entropy ratio by solving Eqs.\ \eqref{rhodot_phi} --
\eqref{rhodot_dark} from the epoch of $\phi\sim O(M_{\rm Pl})$ to the
epoch enough after the reheating.

For the calculation of the mass density of the dark glueball, we need
to estimate when the chemical equilibrium of the dark sector breaks
down and the number of the dark gluon in the comoving volume freezes
out; the cosmic time at the freeze out is denoted as $t_{\rm F}$.  We
evaluate $t_{\rm F}$ by solving the following equation:\footnote
{We use Eq.\ \eqref{FreezeOut} to estimate the freeze-out temperature
  of the dark glueball.  After the freeze out, the inverse process
  becomes irrelevant so that the Boltzmann equation becomes
  \begin{align*}
    \dot{n}_X + 3 H n_X \simeq - \frac{c_{32}}{\Lambda^5} n_X^3.
  \end{align*}
  Assuming the radiation dominance at the time of the freeze out, the
  evolution of the number density after the freeze out is given by
  \begin{align*}
    n_X (t \gg f_F) \simeq 
    n_X (f_F)
    \left[
      1 + \frac{\Gamma_{3X\leftrightarrow 2X}(t_{\rm F})}{2H(t_{\rm F})}
      \right]^{-1/2}
    \left[ \frac{a(t_{\rm F})}{a(t)} \right]^3.
  \end{align*}
  Thus, the number of the glueball in the comoving volume is almost
  conserved if $H\gtrsim \Gamma_{3X\leftrightarrow 2X}$.  In
  \cite{Carlson:1992fn}, the freeze-out condition is slightly
  different; we have checked that the mass density of the glueball does
  not change significantly even if we use the freeze-out condition
  given in \cite{Carlson:1992fn}.}
\begin{align}
  H(t_{\rm F}) = \Gamma_{3X\leftrightarrow 2X} (t_{\rm F}),
  \label{FreezeOut}
\end{align}
where the right-hand side is the scattering rate for the
$3X\leftrightarrow 2X$ scattering processes in the chemical
equilibrium.  The accurate calculation of the scattering rate is
hardly performed because of the strongness of the interaction.  Here,
from a dimensional consideration, we parameterize
\begin{align}
  \Gamma_{3X\leftrightarrow 2X} = c_{32}
  \frac{n_X^2}{\Lambda^5},
\end{align}
where $c_{32}$ is a constant; we assume $c_{32}\sim O(1)$.  In fact,
the resultant mass density of the dark glueball is insensitive to the
parameter $c_{32}$ because the number density of the dark glueball
approximately scales as $a^{-3}$ (although, as mentioned before, it
decreases slightly faster than $a^{-3}$).  Because
$\Gamma_{3X\leftrightarrow 2X}$ has an exponential dependence on
$T_{\rm dark}$, the dark-sector temperature at the time of the freeze
out is insensitive to $\xi_h$.  Let us define 
\begin{align}
  x_{\rm F} \equiv \frac{m_X}{T_{\rm dark} (t_{\rm F})}.
\end{align}
Then,  with numerically solving
Eq.\ \eqref{FreezeOut}, we found $x_{\rm F}\sim 20$ 
for $m_X=100\ {\rm MeV}$.

\begin{figure}[t]
  \centering
  \includegraphics[width=0.75\textwidth]{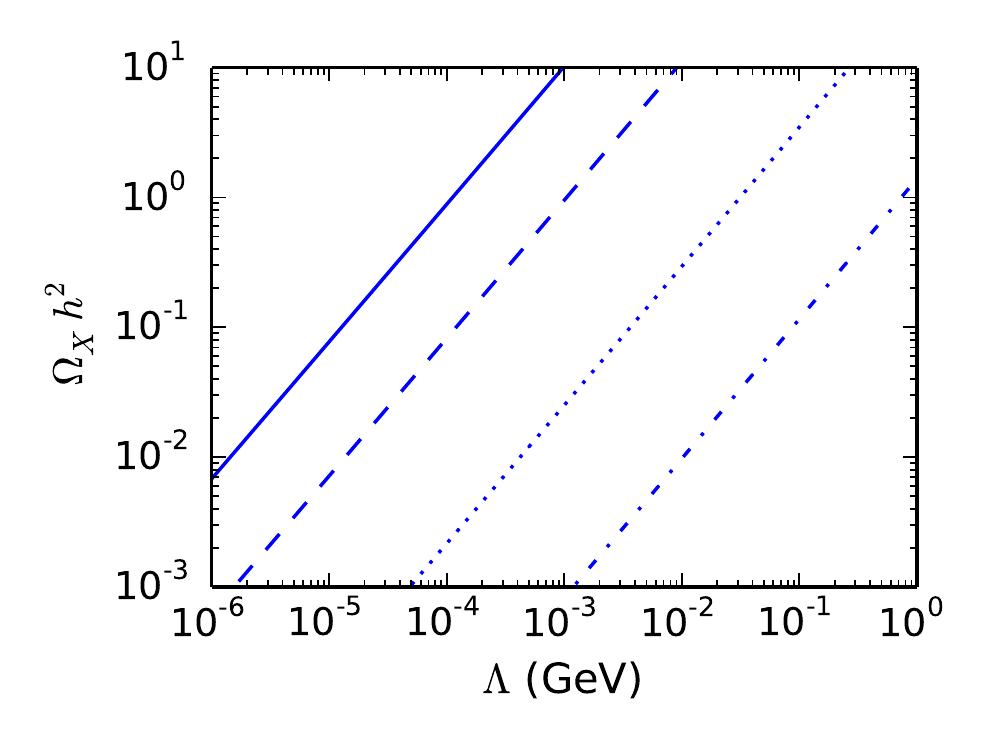}
  \caption{The density parameter of the dark glueball, $\Omega_Xh^2$,
    as a function of $\Lambda$, taking
    $m_X=\Lambda$ and $c_{32}=1$.  The non-minimal coupling of the
    Higgs boson is taken to be $\xi_h=0$ (solid), $1$ (dashed),
    $10$ (dotted), and $100$ (dash-dotted).}
  \label{fig:omghh}
\end{figure}

We estimate the present mass density of the dark glueball as
\begin{align}
  \frac{\rho_X}{s_{\rm vis}} \sim
  \left. \frac{s_{\rm dark}}{s_{\rm vis}}
  \frac{m_X n_X}{s_{\rm dark}} \right|_{t=t_{\rm F}}
  \sim 
  \frac{s_{\rm dark}}{s_{\rm vis}} \frac{m_X}{x_{\rm F}},
  \label{rhox/s}
\end{align}
where, in the second equality, we have used the relation $n_X/s_{\rm
  dark}\simeq T_{\rm dark}/m_X$ in the non-relativistic limit.  The
ratio of the entropy densities $s_{\rm dark}/s_{\rm vis}$, which is
independent of time, is calculated by numerically solving
Eqs.\ \eqref{rhodot_phi} -- \eqref{rhodot_dark}, as we have mentioned.
The ratio is insensitive to $\Lambda$ because it is determined by the
relative size of $\Gamma_{\phi\rightarrow {\rm vis}}$ and
$\Gamma_{\phi\rightarrow {\rm dark}}$; for $\Lambda\sim 100 {\rm
  MeV}$, we found
\begin{align}
  \frac{s_{\rm dark}}{s_{\rm vis}} \sim
  1 \times 10^{-3}
  \times |1 - 6\xi_h|^{-3/2}.
\end{align}
Comparing Eq.\ \eqref{rhox/s} with $\rho_{\rm crit}/s_{\rm now}\simeq
3.6 h^2 \times 10^{-9}\ {\rm GeV}$ (where $\rho_{\rm crit}$ and
$s_{\rm now}$ are critical density and the present entropy density,
respectively, and $h$ is the present expansion rate in units of
$100\ {\rm km/sec/Mpc}$), we can obtain the density parameter of the
dark gluon: $\Omega_X = m_X n_X(t_{\rm now})/\rho_{\rm crit}$ (with
$t_{\rm now}$ being the present cosmic time).  In
Fig.\ \ref{fig:omghh}, we show $\Omega_Xh^2$ as a function of
$\Lambda$ for several values of $\xi_h$; the density parameter is
approximately obtained as
\begin{align}
  \Omega_Xh^2 \sim 
  2\times 10^3
  \times
  |1 - 6\xi_h|^{-3/2}
  \left( \frac{\Lambda}{100\ {\rm MeV}} \right),
\end{align}
where we have assumed that the inflaton dominantly decays into the SM
Higgs pair.

\subsection{Constraints}

For the dark glueball DM, we should consider several
cosmological and astrophysical constraints.  Hereafter, the constraints
are discussed in order.

The dark glueball may have sizable velocity at the time of the freeze
out, and it may affect the structure formation.  In particular, if the
glueball mass is too small, it may affect the formation of the damped
Lyman-$\alpha$ system; according to \cite{deLaix:1995vi}, the dark
glueball mass as the dominant component of DM is required to
be heavier than $\sim O(10)\ {\rm keV}$, and hence 
\begin{align}
  \Lambda \gtrsim O(10)\ {\rm keV}.
\end{align}

From the observations of bullet clusters, the self interaction of the
DM is known to be weak. The self-interaction cross section
$\sigma_{2X\leftrightarrow 2X}$ is bounded from above
\cite{Clowe:2003tk, Markevitch:2003at, Randall:2007ph, Harvey:2015hha,
  Robertson:2016xjh}; here we adopt
\begin{align}
  \frac{\sigma_{2X\leftrightarrow 2X}}{m_X} \lesssim 1 \ {\rm cm^2/g},
\end{align}
Assuming $\sigma_{2X\leftrightarrow 2X}\sim\Lambda^{-2}$ based on the
dimensional analysis, we obtain
\begin{align}
  \Lambda \gtrsim 60\ {\rm MeV}.
  \label{bulletclusters}
\end{align}
Then, it requires $\xi_h\gtrsim 70$ to make the dark gluon density
consistent with the observed DM density of $\Omega_{\rm
  DM}h^2\simeq 0.12$~\cite{Aghanim:2018eyx}.  Such a value of $\xi_h$
is, however, disfavored from the point of view of the stability of the
electroweak vacuum (see below).

We should also consider the stability of the electroweak vacuum during
and after inflation.  It has been well known that the electroweak
vacuum is unstable if the SM is valid up to a scale much higher than
the electroweak scale (like the Planck scale).  With the central
values of the SM parameters inferred from experiments, the lifetime of
the electroweak vacuum for the flat spacetime has been calculated to
be much longer than the present cosmic time \cite{Degrassi:2012ry,
  Andreassen:2017rzq, Chigusa:2017dux, Chigusa:2018uuj}.  However, the
stability of the electroweak vacuum is not guaranteed
during~\cite{Kobakhidze:2013tn, Fairbairn:2014zia, Hook:2014uia,
  Herranen:2014cua, Kamada:2014ufa, Kearney:2015vba, Espinosa:2015qea,
  Mantziris:2020rzh} and after inflation~\cite{Herranen:2015ima,
  Ema:2016kpf, Kohri:2016wof, Enqvist:2016mqj, Postma:2017hbk,
  Ema:2017loe, Ema:2017rkk, Figueroa:2017slm, Rusak:2018kel} in
particular with the non-minimal coupling of the SM Higgs doublet.  In
the present scenario, the inflaton is assumed to decay into the SM
Higgs field via the following interaction:
\begin{align}
  {\cal L} \ni \xi_h \hat{R} h^\dagger h.
  \label{Rhh}
\end{align}
In the inflaton oscillating era after inflation, the non-minimal
coupling of the Higgs doublet is dangerous because the oscillating
behaviors of $R$ (as well as that of the inflaton) may cause an
instability \cite{Ema:2016kpf}.  
The Einstein frame action of the Higgs is given by
\begin{align}
	S = \int d^4x \sqrt{-g}\left[ \varphi^{-1} g^{\mu\nu}(D_\mu h)^\dagger (D_\nu h) 
	+ \xi_h \varphi^{-2} \hat{R}\,h^\dagger h \right].
\end{align}
By defining the canonical Higgs field as $\varphi^{-1/2}h$, the effective mass of the canonical Higgs field is calculated as\footnote
{In conventional inflation models like the chaotic inflation model,
  the expression of the Higgs effective mass is different and is given
  by $\Delta m_h^2 \equiv - \xi_h R$.}
\begin{align}
  \Delta m_h^2 = -\xi_h \varphi^{-1} \hat R - \frac{1}{\sqrt 6 M_{\rm Pl}} \frac{\partial V}{\partial \phi}-\frac{\dot\phi^2}{6M_{\rm Pl}^2}.
\end{align}
When the universe is dominated by the inflaton, $\hat{R}$ is given by
\begin{align}
  \hat{R} = \exp \left( \sqrt{\frac{2}{3}} \frac{\phi}{M_{\rm Pl}}\right)
  \left[ 
    R
    - \frac{\sqrt{6}}{M_{\rm Pl}} \frac{\partial V}{\partial \phi}
    - \frac{\dot{\phi}^2}{M_{\rm Pl}^2}
    \right],
    \label{Ricci_Jordan}
\end{align}
with
\begin{align}
  R = - \frac{1}{M_{\rm Pl}^2} (4V - \dot{\phi}^2).
\end{align}
During inflation, we obtain $\Delta m_h^2 \simeq 12\xi_h H^2$.
Thus, if $\xi_h$ is positive and is larger than $\sim 0.1$, the Higgs
field is forced to stay at the origin during the inflation even with
taking into account the quantum fluctuation; it can help stabilizing
the electroweak vacuum.  In \cite{Mantziris:2020rzh}, a lower bound
$\xi_h \gtrsim 0.05$ was obtained in the Starobinsky inflation.  In
the inflaton-oscillating era after inflation, on the contrary, the
non-minimal coupling of the Higgs doublet may destabilize the
electroweak vacuum because $\Delta m_h^2$ oscillates and it takes
negative value during the oscillation.  The constraint on the Higgs
non-minimal coupling has been studied in \cite{Ema:2016kpf} in the
chaotic inflation model.  We translate the result of
\cite{Ema:2016kpf} to constrain the non-minimal coupling of the Higgs
in the Starobinsky inflation model.  (More detailed study about the
stability of the electroweak vacuum in the Starobinsky inflation is
left as a future project \cite{LiMoroiNakayamaYin_Future}.)  For the
chaotic inflation, the smallest effective mass allowed by the
stability of the electroweak vacuum is found to be about $-5\times
10^{27}\ {\rm GeV}^2$.  Using the fact that the maximal value of
$\varphi^{-2}\hat{R}$ in the Starobinsky inflation model is about
$9\times 10^{26}\ {\rm GeV}^2$, we find the upper bound on the
non-minimal coupling to be\footnote{ Note that the second term of
  (\ref{Ricci_Jordan}), which is linear in $\phi$, is the dominant
  term to derive this bound. The same term also induces the inflaton
  perturbative decay into the Higgs pair.  }
\begin{align}
  \xi_h \lesssim 6.
  \label{xi<6}
\end{align}

The constraint \eqref{xi<6} from the stability of the electroweak
vacuum is in tension with the lower bound of $\xi_h\gtrsim 70$ from
the bullet cluster constraint \eqref{bulletclusters}.  We have come to
this conclusion with the study of the case of $N=2$.  For the case of
$N\geq 3$, as we have mentioned, the confinement phase transition is
expected to be first order \cite{Lucini:2005vg}.  Thus, in such a
case, an extra entropy production in the dark sector is expected as a
result of the release of the latent heat, which enhances the energy
density of the dark sector compared to the case of second-order phase
transition.  Thus, for $N\geq 3$, the value of $\xi_h$ to realize
$\Omega_X=\Omega_{\rm DM}$ becomes larger for a fixed value of
$\Lambda$.  As a consequence, the tension in two constraints is
expected to become  more severe for $N\geq 3$.  Thus, in the minimal setup,
the dark glueball can hardly play the role of DM with satisfying
astrophysical and cosmological constraints.  In the next Section, we
discuss possibilities to make the dark glueball a viable candidate of
DM.

Before closing this section, we comment on the stability of the dark
glueball.  In the present scenario, the stability is not guaranteed by
any symmetry \cite{Gross:2020zam}.  The dark glueball may decay into a
pair of the SM particles if there exists a direct interaction between
the dark sector and the SM sector.  Allowing Planck-suppressed
higher dimensional operators connecting two sectors, the following
dimension-$6$ operator determines the lifetime of the dark glueball:
\begin{align}
  {\cal L}_{\rm int} = \frac{c_{hhGG}}{M_{\rm Pl}^2}
  h^\dagger h
  \mathcal{G}_{\mu\nu}^{(a)} \mathcal{G}^{(a)\mu\nu},
  \label{D6int}
\end{align}
where $\mathcal{G}_{\mu\nu}^{(a)}$ is the field strength of the dark
gluon and $c_{hhGG}$ is a coefficient of $O(1)$.  Assuming that the
decay mode $X\rightarrow\bar{f}f$ (with $f$ being an SM fermion) is
kinematically allowed, the decay rate of the dark glueball is estimated
as $\Gamma_X\sim m_f^2 \Lambda^7/M_{\rm Pl}^4m_h^4$, where $m_f$
denotes the mass of $f$ and $m_h\simeq 125\ {\rm GeV}$ is the Higgs
boson mass. Then, the lifetime is given by
\begin{align}
  \Gamma_X^{-1} \sim 
  O(10^{67})\ {\rm sec} \times
  \left( \frac{m_f}{\Lambda} \right)^{-2}
  \left( \frac{\Lambda}{100\ {\rm MeV}} \right)^{-9}.
\end{align}
Thus, for the mass range of the dark glueball considered in our
analysis, the lifetime is much longer than the present cosmic time and
we can safely neglect the constraints on decaying DM.  Notice that, if
the dark glueball is heavier, its lifetime becomes shorter.  In
particular, if the dark glueball is heavier than the electroweak scale,
it is severely constrained by astrophysical and cosmological
observations.  Unstable DM decaying into the SM particles with $1\ {\rm
  sec}\lesssim \Gamma_X^{-1}\lesssim 10^{26}\ {\rm sec}$ affects the
big-bang nucleosynthesis, the spectrum of the CMB, and/or the flux of
high energy cosmic rays and hence is severely constrained.  Such
constraints excludes the dynamical scale of $10^{5}\ {\rm GeV}\lesssim
\Lambda\lesssim 10^{10}\ {\rm GeV}$ \cite{Gross:2020zam}.

\section{Dark Glueball as Dark Matter}
\label{sec:glueballdm}
\setcounter{equation}{0}

In the previous section, we have seen that the mass density of the
dark glueball hardly becomes consistent with the present DM density in
the minimal case, adopting the astrophysical and cosmological
constraints.  In the parameter region consistent with the constraints,
the relic abundance of the dark glueball becomes so large that the
dark glueball cannot play the role of DM.  In non-minimal cases, on
the contrary, this conclusion may be altered.  One possibility to
realize the dark glueball DM is to introduce an extra entropy
production in the SM sector after the reheating due to the inflaton
decay.  With an extra entropy production, the dark glueball density is
diluted so that the difficulty due to the overproduction may be
avoided.  Another possibility is to add extra fields to make the electroweak vacuum
(absolutely) stable or to enhance the inflaton decay rate into the visible sector. 
In the following, we consider such possibilities.

\subsection{Entropy production}

The extra entropy production often occurs when there exists a
long-lived particle which dominates the universe before it decays.  In
the minimal setup discussed in the previous section, there is no
candidate of such a long-lived particle.  Here, we consider a simple
extension of the setup; we introduce a new sector containing another
non-Abelian gauge interaction.  For simplicity, we assume that the new
sector, called dark$'$ sector, consists only of $SU(N')$ gauge boson
with its dynamical scale $\Lambda'$.  (Here and hereafter, we put
``prime'' on quantities in the dark$'$ sector.)  Then, because the
inflaton of the Starobinsky inflation couples universally to the trace
of the energy momentum tensor, it decays also into the dark$'$ gauge
bosons.  With the confinement of the $SU(N')$ gauge interaction,
glueball in the dark$'$ sector (which we call $X'$) should be formed
and it behaves as non-relativistic matter.  Assuming an interaction
connecting dark$'$ sector and the SM sector, like the one given in
Eq.\ \eqref{D6int}, $X'$ decays into SM particles.  If the dynamical
scale $\Lambda'$ is high enough, the lifetime of $X'$ becomes shorter
than the present cosmic time.  In particular, with a proper choice of
$\Lambda'$, $X'$ dominates the universe before its decay, and an extra
entropy production may occur which reduces the mass density of $X$.

In the following, we quantitatively estimate the dilution factor due
to the decay of $X'$ and see if the mass density of $X$ can become
consistent with the present DM density.  For simplicity, we
concentrate on the case with $N'=2$ and $\Lambda'\sim 10^{12}\ {\rm GeV}$.
Notice that the dynamical scale of $SU(N')$ is assumed to be lower
than the inflaton mass so that the inflaton can decay into the dark$'$
gauge bosons; the decay rate is given in Eq.\ \eqref{Gamma(gauge)}.
In addition, in order to make $X'$ decay into SM particles, we assume
that there exists the following operator:
\begin{align}
  {\cal L}_{\rm int} \ni \frac{1}{{M'}^2}
  h^\dagger h
  {\mathcal{G}'}_{\mu\nu}^{(a)} {\mathcal{G}'}^{(a)\mu\nu},
  \label{D6int'}
\end{align}
where ${\mathcal{G}'}_{\mu\nu}^{(a)}$ is the field strength of the
$SU(N')$ gauge field.  The decay rate of $X'$ is estimated as
\begin{align}
  \Gamma_{X'\rightarrow h^\dagger h} \sim 
  \frac{\Lambda'^5}{{M'}^4}.
\end{align}
The ``cut-off'' scale ${M'}$ is model-dependent.  The interaction may
arise from the effect of quantum gravity with ${M'}\sim O(M_{\rm
  Pl})$.  A smaller value of the cut-off scale is also possible in
particular if there exists a heavy particle, which couples to the SM
Higgs doublet, with an $SU(N')$ quantum number.  An example is an
$SU(N')$ non-singlet heavy scalar particle (which we call $\eta$) with
the quartic interaction of $\sim h^\dagger h \eta^\dagger \eta$; in
such a case, ${M'}$ is given by the mass scale of $\eta$ (with a loop
factor).  We do not specify the origin of the interaction given in
Eq.\ \eqref{D6int'} and leave ${M'}$ as a free parameter.

Let us estimate the dilution factor due to the decay of $X'$.  As in
the case discussed in the previous section, the decay process into the
dark$'$ sector occurs via the trace anomaly.  Because we take
$\Lambda'\gg\Lambda$, $\alpha_{\rm dark}' (m_\phi)$ (which is the
$SU(N')$ gauge coupling constant at $Q=m_\phi$) is  larger than
$\alpha_{\rm dark} (m_\phi)$.  Consequently, the partial decay rate
$\Gamma_{\phi\rightarrow{\rm dark}'}$ becomes significantly larger
than $\Gamma_{\phi\rightarrow{\rm dark}}$, and hence the production of
the $SU(N')$ gauge boson is more effective than that of $SU(N)$ gauge
boson because
\begin{align}
  \frac{\rho_{\rm dark} (t_{\rm R})}{\rho_{{\rm dark}'} (t_{\rm R})} \sim
  \frac{\Gamma_{\phi\rightarrow{\rm dark}}}{\Gamma_{\phi\rightarrow{\rm dark}'}}.
\end{align}
With our choice of parameters,
$\Gamma_{\phi\rightarrow{\rm dark}}/\Gamma_{\phi\rightarrow{\rm
    dark}'}\sim O(10^{-2})$.

For $\Lambda'\sim 10^{12}\ {\rm GeV}$, (i) the energy density of the
dark$'$ sector is much smaller than $\Lambda'^4$ at the time of the
reheating and hence the dark$'$ sector is always in the confinement
phase after the reheating, and (ii) the scattering rate for the number
changing processes of $X'$ is smaller than the expansion rate.  As a
consequence, $X'$ behaves as a non-relativistic matter after the
reheating (but before the epoch of $H\sim \Gamma_{X'\rightarrow
  h^\dagger h}$), and its energy density scales as $\sim a^{-3}$.
Assuming that $X'$ decays after dominating the universe, the energy
density of $X'$ at the time of the decay (denoted as $t_{\rm dec}$) is
given by $\rho_{X'}(t_{\rm dec})\sim M_{\rm Pl}^2\Gamma_{X'\rightarrow
  h^\dagger h}^2$, and the visible sector temperature after the decay
is estimated as
\begin{align}
  T_{\rm vis} (t_{\rm dec}) \sim 
  \left(\frac{g_\star \pi^2 }{90}\right)^{-1/4}
  \sqrt{ M_{\rm Pl}\Gamma_{X'\rightarrow h^\dagger h}}.
\end{align}
Numerically, 
\begin{align}
  T_{\rm vis} (t_{\rm dec}) \sim 
  O(10^2)\ {\rm GeV} \times
  \left( \frac{g_\star}{106.75} \right)^{-1/4}
  \left( \frac{\Lambda'}{10^{12}\ {\rm GeV}} \right)^{5/2}
  \left( \frac{M'}{M_{\rm Pl}} \right)^{-2}.
\end{align}

Let us define the dilution factor:
\begin{align}
  D \equiv 
  \left. 
  \frac{s_{\rm vis}(t)}{s_{\rm vis}^{({\rm would\mathhyphen be})}(t)}
  \right|_{t\gg t_{\rm dec}},
\end{align}
where $s_{\rm vis}^{({\rm would\mathhyphen be})}$ is the
visible-sector entropy density without the effect of the entropy
production and is given by $s_{\rm vis}^{({\rm would\mathhyphen
    be})}(t)\sim s_{\rm vis}(t_{\rm R}) \times [a(t_{\rm R})/a(t)]^3$.
Using the fact that $s_{\rm vis}^{({\rm would\mathhyphen be})}$ and
$\rho_{X'}$ both scale as $a^{-3}$ for $t_{\rm R}\lesssim t\lesssim
t_{\rm dec}$, the dilution factor is given by
\begin{align}
  D \sim
  \frac{\Gamma_{\phi\rightarrow{\rm dark}'}}
       {\sqrt{\Gamma_{\phi\rightarrow{\rm vis}} \Gamma_{X'\rightarrow h^\dagger h}}}.
\end{align}
Here, it is assumed that $\Gamma_{\phi\rightarrow{\rm
    dark}'}\ll\Gamma_{\phi\rightarrow{\rm vis}}$, and also that
$g_\star (t_{\rm dec})=g_\star (t_{\rm R})$.  With our choice of model
parameters,
\begin{align}
  D \sim O(10^6) \times
  |1 - 6\xi_h|^{-1}
  \left( \frac{\Lambda'}{10^{12}\ {\rm GeV}} \right)^{-5/2}
  \left( \frac{M'}{M_{\rm Pl}} \right)^{2}.
\end{align}
Thus, if $M'$ and $\Lambda'$ are properly chosen, the mass density of
$X$ can be reduced down to the present DM density in the
parameter region consistent with the astrophysical and cosmological
constraints.

So far, we have considered the entropy production due to the decay of
heavy dark glueball.  We comment that the entropy production is also
possible by the decay of visible sector particles.  If there exists an
extra boson or fermion in the visible sector and also if it dominates
the universe at some epoch, its decay into the SM particles dilutes
the dark glueball.  Then, the dynamical scale relevant for the dark
glueball DM can be larger.  One candidate may be the right-handed
neutrino that couples very weakly to the SM leptons; it does not
contribute too much to the active neutrino masses as far as its Yukawa
interaction is weak enough.  If its mass is around $10^{12}$\,GeV, the
production due to the inflaton decay can be efficient.  Then, if the
Yukawa coupling is smaller than $O(10^{-10})$, the mass of the
glueball DM can be heavier than $O(100)$\,MeV.

\subsection{Other loopholes}

Before closing this section, we discuss other possibilities to realize
the dark glueball DM without conflicting the astrophysical and
cosmological constraints.

The main reason for the minimal dark glueball DM scenario does not
work was that the large non-minimal coupling $\xi_h$, required for
correct cold dark glueball DM abundance, would lead to the collapse of
the electroweak vacuum during reheating, as discussed in the previous
section.  However, the stability of the electroweak vacuum crucially
depends on the existence of new particles.

For example, one may introduce an additional scalar field $\Psi$,
which develops a vacuum expectation value $\left<\Psi\right>$ and has
a portal coupling to the SM Higgs as $\sim |h|^2|\Psi|^2$, can make
the electroweak vacuum absolutely stable due to the scalar threshold
effect depending on the mass and coupling constant of the
scalar~\cite{Lebedev:2012zw,EliasMiro:2012ay}. In this case, large
non-minimal coupling $\xi_h$ does not lead to any cosmological
problem. Such a scalar field $\Psi$ may be identified with the $B-L$
Higgs field or the Peccei-Quinn scalar, and hence is theoretically
motivated well.  See also Ref.~\cite{Ema:2016ehh} for other models to
make the electroweak vacuum stable.

If there exist extra scalar fields in association with, for example,
the $B-L$ symmetry or the Peccei-Quinn symmetry, the inflaton may also
decay into those scalars in particular when their non-minimal coupling
$\xi_{\rm extra}$ is sizable.  This fact motivates us to consider
another possibility to realize the dark glueball DM.  If $\xi_{\rm
  extra}$ is sizable, the inflaton may dominantly decay into such an
extra scalar field.  Assuming that the potential of such a scalar
field does not have instability, $\Gamma_{\phi\rightarrow{\rm vis}}$
can be safely enhanced by taking large $\xi_{\rm extra}$.  With large
enough $\xi_{\rm extra}$, the branching ratio into the dark sector is
suppressed and $\Lambda$ consistent with $\Omega_X=\Omega_{\rm DM}$
becomes so large that the bullet cluster constraint can be avoided.

\section{Other Dark Matter Candidates}
\label{sec:others}
\setcounter{equation}{0}

So far, we have studied the possibility of dark glueball DM.  In
this section, we discuss other possibilities, i.e., the cases that the
dark sector contains a scalar field, a fermion, or a massive Abelian
gauge boson.  (They are collectively denoted by $X$ in this Section.)  We assume
that they are so weakly interacting that they freely propagate after
being produced.

\subsection{Dark matter from inflaton decay}

The number density of $X$ (denoted as $n_X$) evolves as
\begin{align}
  \dot{n}_X = 
  - 3 H n_X + \frac{2\Gamma_{\phi\rightarrow {\rm dark}}}{m_\phi} \rho_\phi.
  \label{nxdot}
\end{align}
Here $X$ is assumed to be pair produced by the decay of $\phi$.  We
solve Eqs.\ \eqref{rhodot_phi} and \eqref{rhodot_vis} as well as
Eq.\ \eqref{nxdot} to calculate the relic density of $X$.  Because $X$
survives until today once produced, the number density of $X$ at the
end of the reheating can be approximated as
\begin{align}
  \frac{n_X (t_{\rm R})}{s_{\rm vis} (t_{\rm R})} \sim
  \frac{T_{\rm vis} (t_{\rm R})}{m_\phi}
  \frac{\Gamma_{\phi\rightarrow {\rm dark}}}{\Gamma_{\phi\rightarrow {\rm vis}}}.
\end{align}

For the cases that $X$ is a real scalar, a Dirac Fermion, and a
massive Abelian gauge boson, we calculate the density parameter
$\Omega_X$ as follows.
\begin{itemize}
\item When $X$ is a real scalar field, we can use
  Eq.\ \eqref{Gamma_chichi} to calculate the decay rate.  Assuming that
  the inflaton dominantly decays into the Higgs doublet pair, we find
  \cite{Gorbunov:2010bn,Gorbunov:2012ns}
  \begin{align}
    \Omega_X h^2 \simeq 0.2 \times
    |1 - 6\xi_h|^{-1}|1 - 6\xi_X|^2
    \left( \frac{m_X}{10^{-5}\,{\rm GeV}} \right),
    \label{OmegaX(scalar)}
  \end{align}
  where $\xi_X$ is the non-minimal coupling of $X$.\footnote{
  	In Ref.~\cite{Gorbunov:2012ns} the possibility of scalar DM heavier than the inflaton has also been considered.
	In such a case, the gravitational production is exponentially suppressed but still there may be a parameter region
	that is consistent with the observed DM abundance. We do not pursue this case further in this paper.
  }

\item When $X$ is a Dirac fermion, the partial decay rate of the
  inflaton into the $X$ pair is given in Eq.\ \eqref{Gamma_ff}, and we
  obtain \cite{Gorbunov:2010bn,Gorbunov:2012ns}
  \begin{align}
    \Omega_X h^2 \simeq 0.07 \times
    |1 - 6\xi_h|^{-1}
    \left( \frac{m_X}{10^7\,{\rm GeV}} \right)^3.
  \end{align}

\item When $X$ is a massive Abelian gauge boson, we use
  Eq.\ \eqref{Gamma_VV}.  Then the relic density is given by
  \begin{align}
    \Omega_X h^2 \simeq 0.2\times 
    |1 - 6\xi_h|^{-1}
    \left( \frac{m_X}{10^{-5}\,{\rm GeV}} \right).
  \end{align}

\end{itemize}
For the case that $X$ is a fermion, the relic density consistent with
the present DM density is realized with the mass of $m_X\sim
O(10^7)\ {\rm GeV}$.  On the contrary, for the bosonic case, the mass
of $X$ is required to be as light as $m_X\sim O(10-100)\ {\rm keV}$
for $\Omega_X=\Omega_{\rm DM}$.

\begin{figure}[t]
  \centering
  \includegraphics[width=0.75\textwidth]{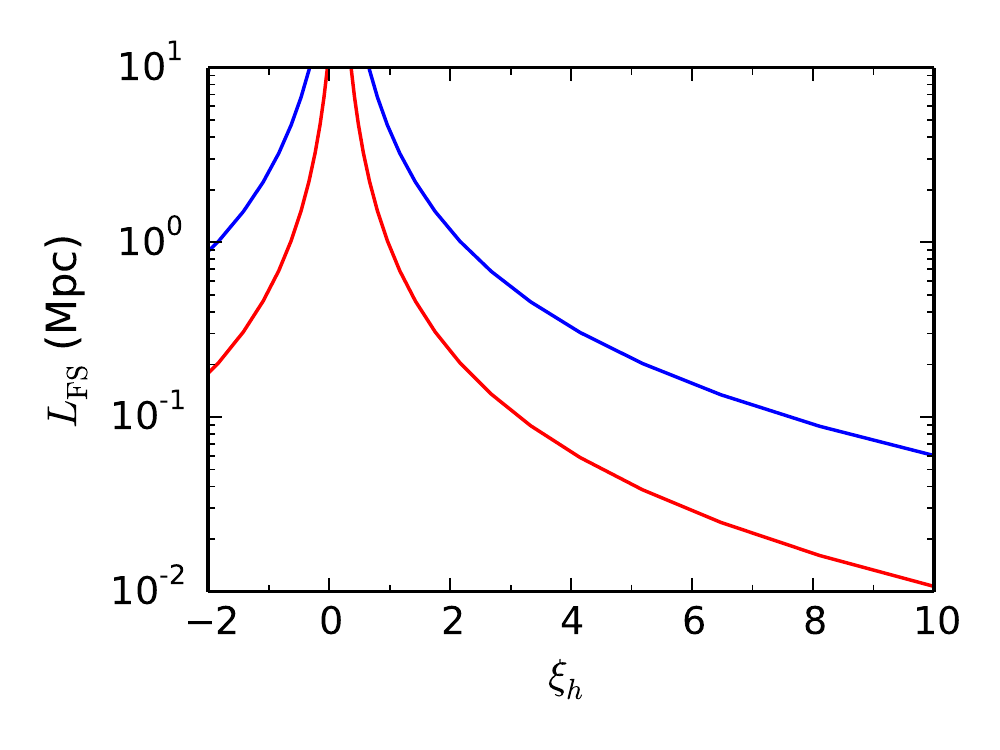}
  \caption{$L_{\rm FS}$ as a function of $\xi_h$ for the case that $X$
    is a scalar boson, taking $\xi_X=0$ (blue) and $0.1$ (red).
    The result for $\xi_X=0$ also applies to the case of
    massive Abelian gauge boson.}
  \label{fig:lfs}
\end{figure}

For light DM candidates, we should consider the constraint from the
Lyman-$\alpha$ forest data \cite{Viel:2005qj, Irsic:2017ixq}.  Such a
constraint is important for the cases of scalar and massive Abelian
gauge boson.  DM models with sizable DM velocity at the time of the
structure formation can be characterized by the free streaming length.
Here, we calculate the averaged free streaming length:
\begin{align}
  L_{\rm FS} \equiv a_{\rm now} \int^{t_{\rm eq}} dt
  \frac{\langle v_X (t) \rangle}{a(t)},
  \label{L_FS}
\end{align}
where $t_{\rm eq}$ is the time of radiation-matter equality and
$a_{\rm now}$ is the scale factor at present.  In addition, $\langle
v_X (t) \rangle$ is the averaged velocity of $X$ produced by the
inflaton decay; in the present case, the energy of $X$ is equal to
$m_\phi/2$ when produced, and the velocity of $X$ is determined solely
by the redshift after the production.  In Fig.\ \ref{fig:lfs}, we plot
$L_{\rm FS}$ as a function of $\xi_h$ for the case that $X$ is a
scalar boson, taking the mass of $X$ realizing $\Omega_X=\Omega_{\rm
  DM}$ (see Eq.\ \eqref{OmegaX(scalar)}).  The non-minimal coupling is
taken to be $\xi_X=0$ and $0.1$.  (Notice that the result for
$\xi_X=0$ also applies to the case of massive Abelian gauge boson.)
The free streaming length becomes smaller in the limit of large
$|1-6\xi_h|$ because, in such a limit, (i) the reheating temperature
is enhanced and hence the velocity of $X$ at the time of the structure
formation is more suppressed due to the redshift, and (ii) the
branching ratio of $\phi$ into $X$ is suppressed so that the mass of
$X$ realizing the correct DM density becomes larger.  The free
streaming length is of $O(10)\ {\rm Mpc}$ for $\xi_h\sim 0$, which is
strongly disfavored, while it can be $O(0.1)\ {\rm Mpc}$ (or smaller)
when $\xi_h$ is larger than a few.  In order to discuss the
consistency with the structure formation, we simply compare $L_{\rm
  FS}$ in the present model with the free streaming length of warm DM
(WDM).  It has been pointed out that the Lyman-$\alpha$ forest
constraint excludes the WDM with its mass lighter than $5.3\ {\rm
  keV}$ \cite{Irsic:2017ixq}, for which the free streaming length
defined in Eq.\ \eqref{L_FS} is $\sim 0.06\ {\rm Mpc}$.  Requiring
$L_{\rm FS}\lesssim 0.06\ {\rm Mpc}$,\footnote
{Rigorously speaking, the velocity distribution of the present DM
  model is different from that in the case of the WDM.  We neglect the
  possible error in the estimation of the upper bound on $L_{\rm FS}$
  due to such a difference.  A more accurate study of the
  Lyman-$\alpha$ forest constraint requires a detailed study with a
  hydrodynamical simulations, which is beyond the scope of our
  analysis.}
the dark scalar field with $\xi_X=0$ hardly becomes a viable
candidate of DM because such a small free streaming length requires
$\xi_h\gtrsim 10$ which conflicts with the constraint from the
stability of the electroweak vacuum (see \eqref{xi<6}).  (The same
conclusion holds for the case of massive Abelian gauge boson.)  With
non-minimal coupling of $\xi_X\sim 0.1$, such a difficulty can be
avoided, as indicated by Fig.\ \ref{fig:lfs}.  In addition, the
minimally coupled dark scalar field and the massive Abelian gauge
boson can be DM without conflicting the astrophysical and cosmological
constraints if there exists a sizable entropy production or if the
electroweak vacuum is somehow stabilized (as we discussed in the case
of dark glueball DM).  We also note here that future 21cm line
observations will improve the sensitivity to the free streaming length
of DM \cite{Sitwell:2013fpa, Sekiguchi:2014wfa}. The Square Kilometer
Array, for example, will have sensitivity to WDM lighter than $\sim
24\ {\rm keV}$ (for which $L_{\rm FS}\lesssim 0.01\ {\rm Mpc}$.  Thus,
the future observations of 21cm line signals will provide crucial test
of the present scenario with the dark scalar DM.

\subsection{Dark matter from scalar coherent oscillation}

For the dark scalar DM $X$, we can consider another possibility than
the production through the inflaton decay, i.e., the misalignment
mechanism~\cite{Preskill:1982cy,Abbott:1982af,Dine:1982ah}. The scalar
amplitude may be non-vanishing during inflation because of the long
wavelength quantum fluctuation or Hubble-induced mass. The former
possibility may be dangerous due to the isocurvature
problem~\cite{Ema:2018ucl}.  In the latter case, the scalar may
acquire a large Hubble-induced mass if non-minimal coupling $\xi_X$ is
$|\xi_X| \gtrsim 1$.

Let us consider the case of $\xi_X<0$. Then $X$ feels the negative
Hubble mass during inflation and its potential minimum is away from
that in the present universe. This may lead to the misalignment
production of DM.  To be concrete, let us assume
\begin{align}
  {\cal L}_X =
  \frac{1}{2} \hat g^{\mu\nu} \partial_\mu X \partial_\nu X
  - \frac{1}{2} m_X^2 X^2- \frac{1}{4}\lambda_X X^4
  + \frac{1}{2}\xi_X \hat R X^2,
\end{align}
with $\xi_X<0$. Here we have introduced a quartic term of $X$.  Then
during inflation, $X$ has the amplitude of
\begin{align}
|X_{\rm inf}| \simeq \frac{ |3\xi_X|^{1/2}\mu}{ \sqrt{\lambda_X}},
\end{align}
due to the valance between the (negative) Hubble induced mass and the
quartic term (see Eq.~(\ref{Ricci_Jordan})). Soon (but not very soon)
after the inflation, the Hubble induced mass becomes negligible since
it is redshifted away faster than the quartic term. Then the $X$
oscillation, driven mainly by the quartic term, starts.  At the end of
the inflation, the oscillation energy density is given by
\begin{align}
  \rho_X (t_{\rm end})
  \sim 
  \frac{\lambda_X X_{\rm inf}^4}{4} \sim
  \frac{9\xi_X^2 \mu^4}{4\lambda_X},
\end{align}
where $t_{\rm end}$ is the cosmic time at the end of the inflation.
The $X$ energy density decreases as that of the radiation as far as
the quartic term is dominant for the coherent oscillation.  Eventually
the mass term becomes dominant and then the $X$ energy density behaves
like that of non-relativistic matter thereafter.  This transition
happens when $m_X^2 X^2/2 \sim \lambda_X X^4/4$; the cosmic time of
the transition is denoted as $t_{\rm tr}$. Then, 
\begin{align}
  \rho_{X} (t_{\rm tr}) \sim \frac{2m_{X}^4}{\lambda_X}.
\end{align}
After this epoch, the ratio $\rho_{X}/s$ becomes a constant and is
estimated as (c.f.~\cite{Daido:2016tsj,
  Daido:2017wwb,Daido:2017tbr,Nakagawa:2020eeg})
\begin{align}
  \frac{\rho_{X}}{s}
  \sim g_\star^{-1/4} (t_{\rm tr})
  |\xi_X|^{3/2} 
  \left( \Gamma_\phi \frac{M_{\rm Pl}^2}{\mu^3} \right)^{1/2}
  \left( \frac{\mu}{M_{\rm Pl}} \right)^{5/2}
  \frac{m_X}{\lambda_X}.
\end{align}
As a consequence, we obtain the DM abundance as
\begin{align}
  \Omega_{X}^{\rm (osc)} h^2 \sim 0.2\times |1-6\xi_h| |\xi_X|^{3/2}  
  \left(\frac{m_X}{100\,{\rm keV}}\right)
  \left(\frac{10^{-8} }{\lambda_X}\right).
\end{align}
We note that $X$ is also produced by the direct decay of the inflaton,
but it is sub-dominant and tends to be hot DM if $m_X \lesssim
O(1)\ {\rm MeV}$. The transition temperature $T_{\rm vis}(t_{\rm
  tr})$ should be larger than $\sim O(1)\ {\rm keV}$. It gives a
lower bound of $m_X$ as $m_X \gtrsim 1\ {\rm eV}$
\cite{Daido:2017wwb,Daido:2017tbr,Nakagawa:2020eeg}.  There is also a
lower bound of $m_X\gtrsim O(1)\, {\rm keV}$ if $|X_{\rm inf}|\lesssim
M_{\rm Pl}$.  Note also that soon after inflation there may be a
tachyonic/parametric resonance if $\xi_X \ll -1$.  On the other hand,
we may need $\xi_X \lesssim -1$ to suppress the isocurvature
bound.\footnote
{This may be
  evaded if $X_{\rm inf}\gtrsim M_{\rm pl}$.}
To clarify the resonance effect on this scenario as well as the
detailed parameter region requires a further study
\cite{LiMoroiNakayamaYin_Future}.

Another simple possibility of realizing the misalignment production is
to (softly) break the $Z_2$ symmetry under which $X$ changes its sign.
Even without the $Z_2$ symmetry, the scalar field is long-lived if it
is light enough while we may introduce the interaction of $\hat R X$.
Neglecting the $X$ self-coupling or the direct coupling to other SM
particles, the Lagrangian is given by\footnote
{In fact, as other studies we have neglected the portal coupling
  between $X$ and SM Higgs field which may be generated in cutoff
  regularization with cutoff scale to be the Planck-scale.  This
  depends on the UV completion of quantum gravity.}
\begin{align}
  {\cal L}_\chi =
  \frac{1}{2} \hat{g}^{\mu\nu} \partial_\mu X \partial_\nu X
  - \frac{1}{2} m_X^2 X^2
  +\xi_X \hat R \left( \frac{1}{2}X^2+ c_X X M_{\rm Pl}\right),
\end{align}
where $c_X$ is the $Z_2$ breaking coupling while the non-minimal
coupling $\xi_X$ is assumed to be positive for simplicity.  We choose
the minimum of the potential at the present universe as $X=0$ without
loss of generality.  Assuming that the bare scalar mass $m_X$ is much
smaller than the expansion rate during inflation, the $X$ amplitude
during inflation is given by $X \simeq -c_X M_{\rm Pl}$ due to the
Hubble induced mass term (with $\xi_X >0$).  After the reheating the
non-minimal coupling becomes irrelevant.  The scalar field $X$ starts
to oscillate when the expansion rate becomes comparable to $m_X$.  The
present energy density of the coherent oscillation is estimated as
\begin{align}
  \Omega_X^{\rm (osc)} h^2\simeq 0.1 \times
  \left(\frac{ |c_X|}{0.01} \right)^{2}
  \left(\frac{m_X}{10^{-19}\,{\rm eV}}\right)^{1/2}. 
\end{align}
Thus the scalar field can be a fuzzy DM.  Assuming $\xi_{X}\gtrsim 1$,
the Hubble-induced mass becomes comparable to or larger than the
expansion rate during inflation.  Then, the quantum fluctuation of
$X$ during inflation is suppressed and hence the isocurvature
problem can be avoided.  The scalar particle $X$ produced by the
inflaton decay is still relativistic in the present universe if
$m_X\lesssim 1\ {\rm eV}$ (for which $c_X\gtrsim 10^7$ to make
$\Omega_X^{\rm (osc)}=\Omega_{\rm DM}$).  In such a case, $X$ produced
by the decay contributes to the dark radiation.  The increase of the
effective neutrino number is estimated as
\begin{align}
  \Delta N_{\rm eff}\simeq  0.8 \times
  \frac{|1-6\xi_X|^2}{|1-6\xi_h|^2}. 
\end{align}
Because $\xi_h \lesssim 6$ due to the vacuum stability bound and also
because $\xi_{X}\gtrsim 1$ for the isocurvature bound, $\Delta
N_{\rm eff}$ is bounded from below as
\begin{align}
  \Delta N_{\rm eff}\gtrsim 0.02.
\end{align}
Thus, interestingly, this scenario can be tested by the future
measurement of dark radiation \cite{Kogut:2011xw,
  Abazajian:2016yjj, Baumann:2017lmt}.

\section{Summary}
\label{sec:conclusions}
\setcounter{equation}{0}

In this paper, we have studied a production mechanism of glueball DM
from a confining dark (pure) $SU(N)$ Yang-Mills theory.  Due to the
gauge symmetry the dark gauge particles are rarely interacting with
the SM particles, and the DM stability is guaranteed if the
confinement scale is much lower than PeV.  This also implies that it
is difficult to be produced thermally if the reheating temperature is
much below the Planck scale.  In the Starobinsky inflation, on the
other hand, it has been argued that the DM is naturally and easily
produced from inflaton decay.  Carefully studying the Higgs potential
stability during and after inflation, the cosmic evolution, and
constraints from bullet clusters as well as structure formations, we
have found that this possibility is disfavored in the minimal setup.
On the contrary, if there is a further entropy dilution to the SM
sector or alternatively the Higgs instability scales happen to be much
higher than central values suggested by the current experimental data,
we can evade the constraints and a dark glueball becomes a viable DM
candidate.  As a concrete example, we build a simple model with an
additional dark pure Yang-Mills gauge sector, whose confinement scale
is close to the inflaton mass.  The decay of this additional glueball
dilutes the DM density to evade the constraints found by us.  To have
the reheating temperature higher than the electroweak scale, the DM
mass is around hundred MeVs and thus this scenario may explain the
small scale crises and alleviate the $S_8$ tension.

We have also discussed the possibility that a scalar, a fermion, or a
massive Abelian gauge boson in the dark sector becomes DM.  As well as
the production of the DM particle from the inflaton decay, we have
discussed the possibility that a scalar field in the dark sector
becomes DM.  If there exists a scalar field in the dark sector, its
coherent oscillation can be induced by the coupling with the inflaton,
which can become a viable candidate of DM.  We show explicit examples
of such scenarios.

The DM candidates in the dark sector interact with SM particles
extremely weakly.  Thus, the direct detection of such a DM is hopeless
with the current technology.  However, with cosmological and
astrophysical observations, we may acquire some hints of the hidden
dark matter in the dark sector.  For example, a signal of the dark
glueball DM may be seen if the observational constraints on the self
interaction of DM can be improved.  In addition, for the light scalar
(or massive Abelian) DM from the inflaton decay, the future 21cm line
observations, which is expected to improve the sensitivity to the free
streaming length of DM, can provide an interesting test.

We have also discussed that the non-minimal coupling of the SM Higgs
plays important roles in the present scenario.  The non-minimal
coupling may enhance the partial decay rate of the inflaton into the
visible sector, while it may be also important to stabilize the
electroweak vacuum during and after inflation.  More detail about this
issue will be discussed in elsewhere \cite{LiMoroiNakayamaYin_Future}.

\vspace{2mm}
\noindent{\it Acknowledgments:} 
This work was supported by JSPS KAKENHI Grant Numbers
16H06490 [TM], 18K03608 [TM], 
17H06359 [KN], 18K03609 [KN],
19H05810 [WY] and 20H05851 [WY].


\bibliographystyle{jhep}
\bibliography{ref}


\end{document}